\documentclass[aps,prl,twocolumn,showpacs,groupedaddress,amsmath,amssymb]{revtex4-1}
\usepackage{graphicx}
\usepackage{dcolumn}
\usepackage{bm}
\usepackage{color}
\usepackage{hyperref}
\usepackage[caption=false]{subfig}

\pdfoutput=1

\hypersetup{colorlinks=true,urlcolor=blue,linkcolor=blue,citecolor=blue}

\newcommand{\bra}[1]{\left\langle#1\right\vert}
\newcommand{\ket}[1]{\left\vert#1\right\rangle}
\newcommand{\sinc}{{\rm sinc}}

\begin{document}

\title{Post-selection free spatial Bell state generation}
\author{E.\,V.\,Kovlakov}\email{ekovlakov@gmail.com} \author{I.\,B.\,Bobrov} \author{S.\,S.\,Straupe} \author{S.\,P.\,Kulik}
\affiliation{Faculty of Physics, M.V.Lomonosov Moscow State
University, Moscow, Russia}

\date{\today}

\begin{abstract}
Spatial states of single photons and spatially entangled photon pairs are becoming an important resource in quantum communication. This additional degree of freedom provides an almost unlimited information capacity, making the development of high-quality sources of spatial entanglement a well-motivated research direction. We report an experimental method for generation of photon pairs in a maximally entangled spatial state. In contrast to existing techniques the method does not require post-selection and allows one to use the full photon flux from the nonlinear crystal, providing a tool for creating high-brightness sources of pure spatially entangled photons. Such sources are a prerequisite for emerging applications in free-space quantum communication.    
\end{abstract}


\maketitle

The process of spontaneous parametric down-conversion (SPDC) is well-known as a tool for quantum engineering of two-photon entangled states. Special attention is paid to maximally entangled Bell-type states which play a crucial role both in fundamental tests of quantum mechanics and practical applications, such as entanglement-based and device-independent quantum cryptography protocols \cite{Ekert,Acin_PRL06}. In this context, the SPDC process is widely utilized as a source of polarization-entangled Bell pairs \cite{kwiat1995}, but over the last decade numerous works have focused on the spatial domain \cite{mair2001,Teich_PRL2007,Padgett_PRA10,Sciarrino_NComm12,mclaren2014}. This interest is driven by the ability to expand the dimensionality of the two-photon Hilbert space using spatial states in addition to polarization and energy-time degrees of freedom \cite{barreiro2005}. 

For example, quantum correlations in the orbital angular momentum (OAM) allow one to produce a Bell pair with a signal photon carrying $\pm l \hbar $ and an idler photon carrying $\mp l \hbar $ units of OAM, respectively \cite{mair2001}. However, existing methods necessarily require postselection  of the states belonging to a particular two-dimensional subspace of a larger Hilbert space. Indeed the two-photon state produced by SPDC with a Gaussian pump may be written as $\ket{\psi} = \sum_{l=-\infty}^{\infty}c_l\ket{l}_s\ket{l}_i$ with $c_l=c_{-l}$. The Bell state is obtained by postselecting a particular value of $l$ \cite{Padgett_OpEx2009,Padgett_PRA10}. The values of $c_l$ are determined by the azimuthal Schmidt number (or spiral bandwidth) of the biphoton state \cite{Torres_PRA2003} and are always less then $c_0$ for $l>0$, so a significant fraction of pairs is lost in the postselection process. An alternative method suggests postselecting a subspace containing a Gaussian $\ket{0}_s\ket{0}_i$ mode \cite{Boyd_PRL2013}, here all four Bell states of the form $\ket{\psi}=\ket{0}_s\ket{0}_i \pm \ket{l}_s\ket{\pm l}_i$ may be generated, but still at the expense of inevitable loss due to postselection. 

A scheme for Bell states generation without the need for postselection was proposed by Yarnall et al. \cite{Teich_PRA2007,Teich_PRL2007}. The authors had mapped the multi-dimensional Hilbert space of transverse modes onto a two-dimensional parity space. Such an approach allowed them to violate the Bell's inequality in the spatial-parity space without subspace projections. This experimental technique is closer to an ideal Bell's experiment, however, the parity states considered are intrinsically multimode, containing as high as 4000 spatial modes \cite{Teich_PRL2007}, complicating their use in free-space communication and almost ruling out the possibility to use them for spatial mode division multiplexing in optical fibers. One should also be very careful not to violate the fair-sampling assumption in Bell experiments with such multimode states \cite{tailored}.  

In this Letter we provide a spatial Bell state generation method, which does not require any spatial filtration of the SPDC transverse mode spectrum. By carefully shaping the spatial profile of the pump we generate the SPDC radiation with the spatial mode content limited to the subspace, containing the zero- and first-order Hermite-Gaussian modes only. High fidelity of the generated states with ideal Bell states is confirmed by full state tomography and significant violation of CHSH inequality. Post-selection free character of the method allows for creation of high-brightness sources of spatial Bell states, while their well-defined and low-order mode content would facilitate their use in free-space and few-mode fiber communication channels. In addition, we study a number of interesting states, generated by transforming the spatial mode of the pump beam to low-order Hermite-Gaussian modes, demonstrating vast abilities in engineering the spatial quantum state of photon pairs.

\emph{Hermite-Gaussian Bell states}. 
The two-photon state generated by SPDC is $\ket{\psi}=\ket{vac} +\mathrm{const}\times\int{\vec{dk_s}\vec{dk_i}\Psi(\vec{k_s},\vec{k_i})\ket{1}_{s}\ket{1}_{i}}$, with $\vec{k}_{s,i}$ being the wave vectors of the scattered photons. Under the paraxial approximation and for the case of the exact collinear phase-matching the biphoton amplitude $\Psi(\vec{k_s},\vec{k_i})$ can be written as
\begin{equation}
\label{Amplitude}
\Psi(\vec{k}_{s \perp},\vec{k}_{i \perp})\propto \mathcal{E}_{p}(\vec{k}_{s \perp}+\vec{k}_{i \perp}) \sinc \Bigr(\dfrac{L(\vec{k}_{s_\bot}-\vec{k}_{i_\bot})^{2}}{4k_{p}} \Bigl),
\end{equation}
where the subscript $\perp$ denotes the transverse vector component. It is clear from this equation that the biphoton spatial wavefunction carries the same functional form as the pump beam amplitude \cite{Burlakov_PRA97,monken1998,walborn2004}, so the control over the spatial shape of the pump beam provides an ability to produce nontrivial two-photon states \cite{Torner_PRL01,Walborn_PRL03,Padgett_JoO12}. 

Let us consider the case in which the nonlinear crystal is pumped with a Hermite-Gaussian (HG) beam, following \cite{Walborn_PRA2005,Guo_PhysLettA05} and \cite{Walborn_JPhysB12}: 
\begin{equation}
	\label{Angular Spectrum}
	\mathcal{E}_{p}^{nm}(k_{x},k_{y},w) =\mathrm{HG}_{n}(k_{x},w) \mathrm{HG}_{m}(k_{y},w),
\end{equation}
where $\mathrm{HG}_{n}(k,w) \propto \sqrt{w} H_{n} ( w k)\exp ( - w^{2} k^{2} /2 )$, $H_{n}(x)$ is the $n$-th order Hermite polynomial, and $w$ is the width parameter. Since HG functions form a complete basis of orthogonal transverse modes, we can decompose the biphoton state as
\begin{equation}
	\label{HG wavefunction}
	\ket{\psi^{(nm)}}= \sum\limits_{j,k,u,t=0}^\infty C_{jkut}^{(nm)} \ket{\mathrm{HG}_{jk}, \sigma_{s}}_{s} \ket{\mathrm{HG}_{ut},\sigma_{i}}_{i}, 
\end{equation}
where $\sigma$ denotes the width of the down-converted mode and kets $\ket{\mathrm{HG}_{\alpha \beta},\sigma}$ are defined as
$
	\label{HG ket}
	\ket{\mathrm{HG}_{\alpha \beta},\sigma} = \int d\vec{k} \mathrm{HG}_{\alpha}(k_{x},\sigma) \mathrm{HG}_{\beta} (k_{y},\sigma) \ket{1}_{\vec{k}}.
$

Now let us model the $\sinc(...)$ function in Eq.(\ref{Amplitude}) with a Gaussian
$ \exp (-\delta^{2} (\vec{k}_{s_\bot}-\vec{k}_{i_\bot})^{2}),$
where  $\delta^{-1} = \sqrt{4k_p/L}$ is the phase-matching angular width \cite{Eberly}.  Under this approximation the two-dimensional wavefunction (\ref{HG wavefunction}) factorizes, such that $C_{jkut}^{(nm)}=C_{ju}^{(n)}C_{kt}^{(m)}$. Furthermore, when the widths of the collected down-converted modes satisfy $\sigma=\sqrt{2 w \delta}$ and the pump beam is a Gaussian one, the HG expansion coefficients take the form $C_{ju}^{(0)}\propto \delta_{ju}$, where $\delta_{ju}$ is the Kronecker delta. This leads directly to a well-known Schmidt decomposition:
\begin{equation}
	\label{Schmidt}
	\ket{\psi^{(00)}}= \sum\limits_{a,b=0}^\infty \sqrt{\lambda_{a} \lambda_{b}} \ket{\mathrm{HG}_{ab},\sigma}_{s} \ket{\mathrm{HG}_{ab},\sigma}_{i},
\end{equation}
experimentally studied in Ref.~\cite{Straupe} (here $\lambda_{n}$ denote the eigenvalues of the corresponding Schmidt modes). The Schmidt number, quantifying the degree of spatial entanglement in this system, becomes factorized: $K=K_{x}\times K_{y}$, where
\begin{equation}
	\label{SchmidtN}
	K_{x,y}=\dfrac{1}{\sum \lambda_{a,b}^{2}}=\dfrac{(w^{2}+\delta^{2})}{2w\delta}.
\end{equation}

Focusing the pump beam such, that $w\approx \delta$ allows one to reach a single spatial mode regime of SPDC, when $K\approx1$. In this case there is only one non-zero eigenvalue in the Schmidt decomposition $\lambda_0=1$, and (\ref{Schmidt}) reduces to a factorized separable state with the signal and idler photons in Gaussian modes. At the same time, if the pump beam is an $\mathrm{HG}_{10}$ mode of the same width $w$, the SPDC state becomes
\begin{equation}
\label{Bell}
\ket{\psi^{(10)}}=\dfrac{\ket{\mathrm{HG}_{00},\mathrm{HG}_{10}}+\ket{\mathrm{HG}_{10},\mathrm{HG}_{00}}}{\sqrt{2}}=\ket{\Psi^{+}},
\end{equation}
which is exactly the Bell state $\ket{\Psi^+}$. Importantly, under the double-Gaussian approximation the amplitudes of the higher-order modes are all exactly equal to zero, and the whole state is localized in the two-dimensional subspace. Thus, the photon pairs are generated in the Bell state by design, without the need to postselect the particular subspace in the measurement process. One may expect, that the Gaussian approximation may be over-optimistic and the exact state will have contributions from higher-order modes. While this is, strictly speaking, true, our experimental results show, that eq. (\ref{Bell}) is a very good approximation to the true state, and the population outside the two-dimensional manifold is insignificant.

\begin{figure}[h]
	\center{\includegraphics[width=0.95\linewidth]{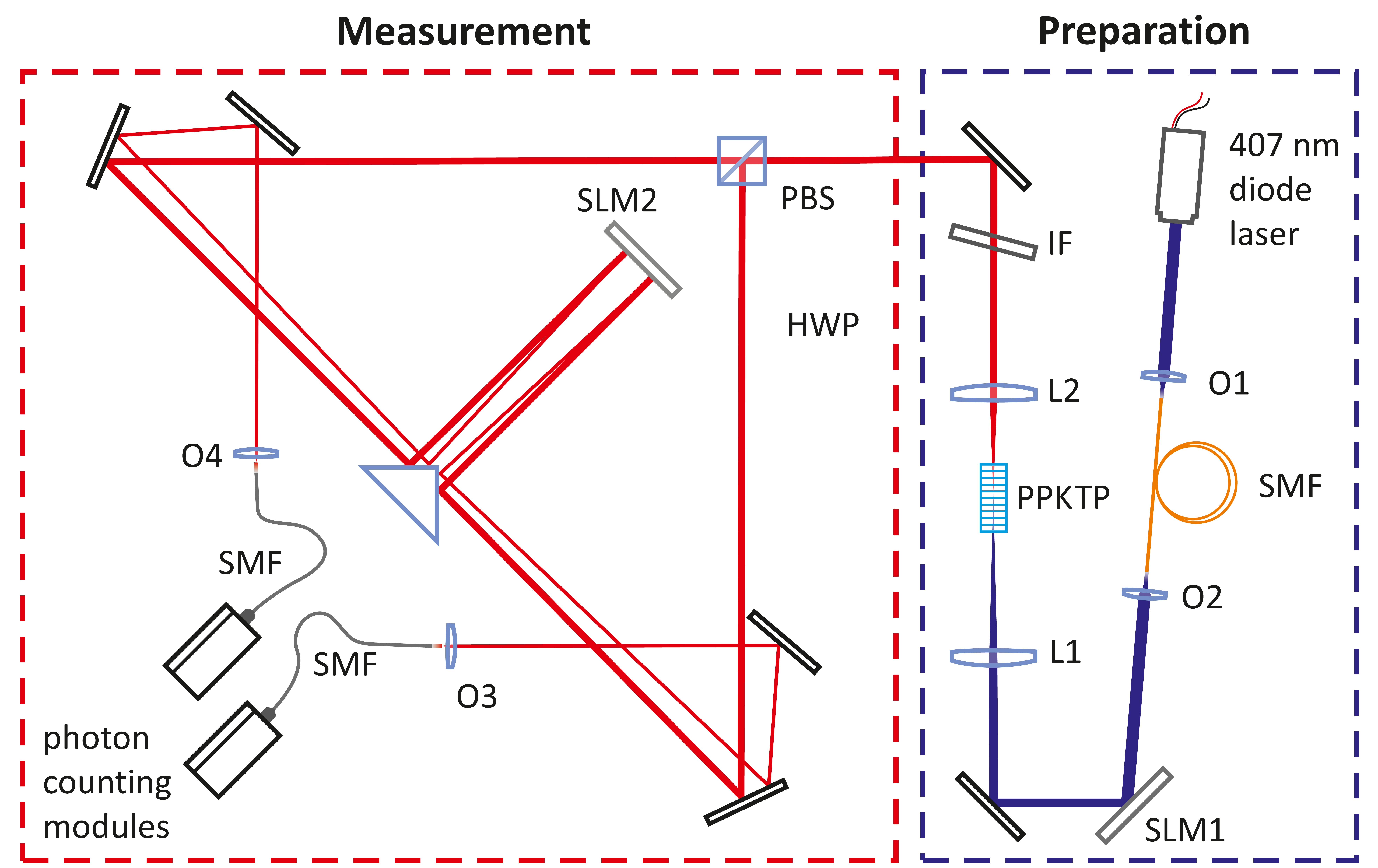}}
	\caption{Experimental setup (see text for a detailed description).}\label{fig:Setup}
\end{figure}

\emph{Experiment.}
The experimental setup used to test the proposed method is shown in Fig.~\ref{fig:Setup}. A 407~nm diode laser was used as a pump. The pump beam was spatially filtered by a single-mode fiber, collimated by a 20X microscope objective (O1), and directed onto a spatial light modulator (SLM1) (Cambridge Correlators). The beam in the first diffraction order of the SLM1 was focused via a 200 mm lens (L1) on a 25-mm thick periodically poled KTP crystal (PPKTP) designed for a collinear frequency degenerate type-II phase-matching. The down-converted light was collimated by a lens (L2) with a focal length of 150 mm and filtered by a 820 nm interference filter (IF) with a 10 nm bandwidth (it was tilted to transmit 814~nm radiation). The produced pairs of vertically and horizontally polarized photons were split on a polarizing beam splitter (PBS)  and reflected by a right angle mirror to the SLM2 (Holoeye Pluto). The right and left halves of the same SLM were used for H- and V-polarized photons, respectively. A half-wave plate (HWP) was installed in the V-channel to match the horizontal working polarization of the SLM2. Photons in the first diffraction order of SLM2 were coupled with 20x microscope objectives (O3) and (O4) to single-mode fibers placed in the objectives focal planes. The fibers were connected to photon counting modules (Perkin-Elmer), followed by a coincidence circuit with a gating time of 4 ns. 

We used a long PPKTP crystal to reach the high SPDC efficiency and to avoid the detrimental walk-off of the interacting beams at the same time. Digital holograms used to generate the pump beam in the desired mode by SLM1 and to detect the resulting SPDC mode spectrum by SLM2 were calculated according to the method proposed in Ref.~\cite{Exact} for phase-only SLMs. This method enables the simultaneous modulation of the phase and amplitude of the beam without unwanted alterations caused by diffraction in the first order. The quality of the projective measurements implemented with this technique have been recently successfully tested experimentally \cite{bent2015,bobrov2015}. 

In the single-Schmidt mode regime the Rayleigh range of the pump beam $z_{R}=\pi w_{p}^{2}/\lambda$ should be equal to the half of the crystal length $L$. Following this criteria, we estimated the required pump beam waist to be $w_{p}=\sqrt{L/k_{p}}\approx 30~ \mu m$ with the down-converted mode waists $\sigma_{s,i}=\sqrt{2 w \delta}=\sqrt{2} w_{p} \approx 43~ \mu m$ 
\footnote{We assume the Gaussian pump beam profile $\mathcal{E}_{p}(x,w_{p}) \propto \exp{(-x^{2}/w_{p}^{2})}$, and define the waist to be $w_p$. We should note, that the width parameter $w$ used in (\ref{SchmidtN}) is the beam waist times $\sqrt{2}$, to comply with the definitions in \cite{Eberly} and the following literature}. With the 150~mm detection lens L2 we obtained $\sigma_{s,i}= 42.6 \pm 1.6~ \mu m$. Mask widths of detection holograms on SLM2 were fixed at the size of detection fiber mode. Adjusting the Gaussian mask width on the SLM1 we were able to maximize the coupling ratio $\eta_{c}=R_{c}/\sqrt{R_{i} R_{s}}$, with $R_{i}$ and $R_{s}$ being the single photon counting rates and $R_{c}$ -- the coincidence rate. The maximum was achieved for the waist $w_{p}= 35.1 \pm 1.1~\mu m$. 

\begin{figure}[h]
	\center{\includegraphics[width=0.95\linewidth]{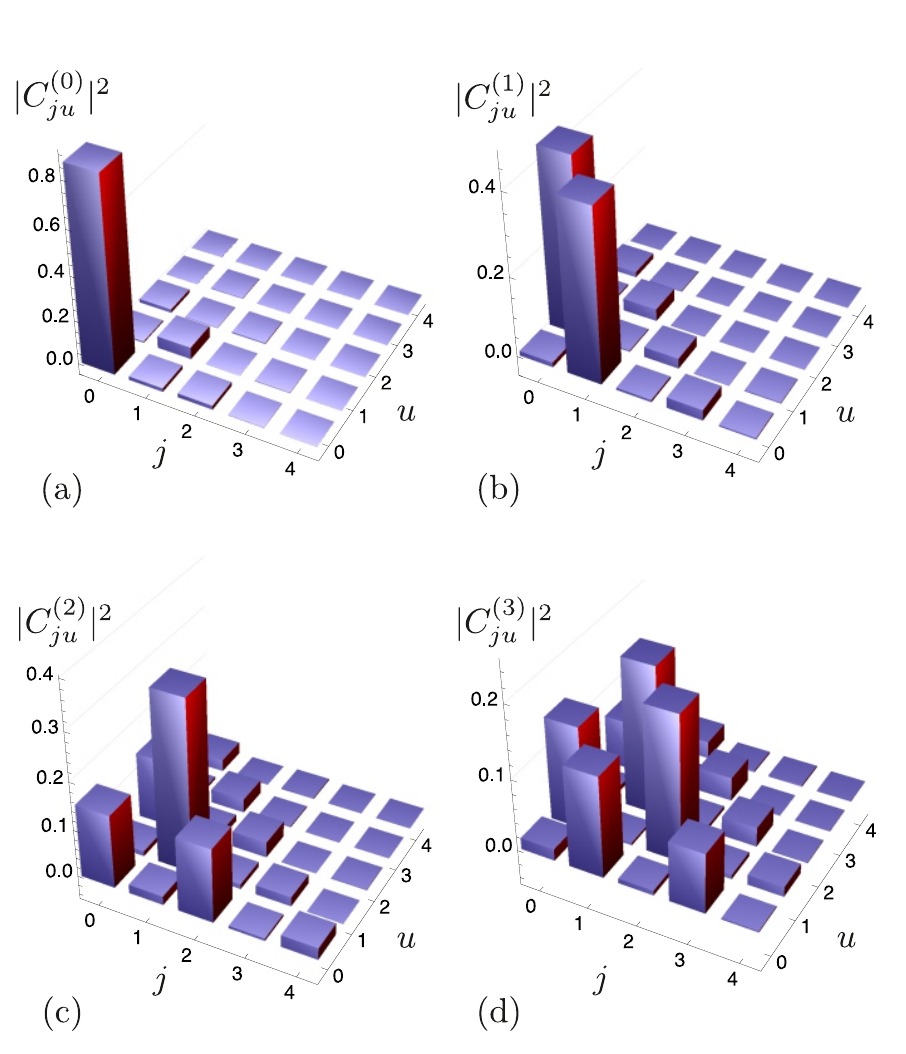}} 
\caption{Experimentally measured coefficients of the mode decomposition for the biphoton amplitude in the Hermite-Gaussian basis (horizontal direction) with the pump beam in the modes: (a) $\mathrm{HG}_{00}$, (b) $\mathrm{HG}_{01}$, (c) $\mathrm{HG}_{02}$ and (d) $\mathrm{HG}_{03}$.
} 
\label{fig:hist}
\end{figure}

Next, we performed Schmidt number measurements with a Gaussian pump beam by projecting the SPDC state on HG modes with SLM2 followed by single mode fibers. The values of  $|C_{ju}^{(0)}|^{2}$ and $|C_{kt}^{(0)}|^{2}$ coefficients for the horizontal and vertical modes were obtained from the experimentally measured coincidence rates for the detection modes $\mathrm{HG}_{jk}$ and $\mathrm{HG}_{ut}$ in the signal and idler channels, respectively. Each histogram was normalized by dividing its elements by the sum of all the elements. One can conclude by inspecting the Fig.~\ref{fig:hist}(a), where a histogram for the horizontal modes is shown, that we are indeed very close to a single-spatial-mode regime of SPDC. The histograms, however, contain barely visible peaks, corresponding to $|C_{02}^{(0)}|^{2}$ and  $|C_{20}^{(0)}|^{2}$, with the height of approximately $2\%$ of the main peak each. Moreover, the variation of the ratio between the pump and the detection waists $w_{p}/\sigma_{s,i}$ in the range from 0.7 to 1.4 did not help us to completely get rid of these higher-order modes \cite{
salakhutdinov2012}. Thus, we attribute the presence of these non-diagonal elements in the matrices $|C_{ju}^{(0)}|^{2}$ and $|C_{kt}^{(0)}|^{2}$ to the notable difference between the real Schmidt basis and the basis of Hermite-Gaussian modes \cite{Miatto_EPJD2012}. This is a consequence of a non-Gaussian nature of the biphotons wavefunction, which makes the exactly single-Schmidt-mode down-conversion regime impossible to realize with a Gaussian pump. Using the relation $\lambda_{a} \propto |C_{aa}^{(0)}|^{2}$ we obtained a robust estimate for the horizontal Schmidt number $K_{x}= 1.31 \pm 0.05$. The similar estimation for the vertical modes gave us $K_{y}= 1.33 \pm 0.04$ with the total Schmidt number $K=K_{x}\times K_{y}=1.74 \pm 0.08$. This was the minimal value of $K$ observed in our experiment (for further details, see Supplementary Materials).

After that, we changed the spatial mode of the pump beam to the first three low-order horizontal HG modes, retaining the optimal focusing conditions. Fig.~\ref{fig:hist} shows the experimentally measured histograms. At the same time we measured the coefficients $|C^{(m)}_{kt}|^{2}$ for the corresponding vertical modes and ensured, that the observed angular spectrum of the vertical modes retains a Gaussian form, confirming the factorization of x- and y-coordinates.  
It is clear from these results that the sum $j+u$ of the mode indices of the down-converted photons is close to, but not exactly equal to the index of the pump mode \cite{Walborn_PRA2005}, due to the fact that $K>1$. 

\begin{figure}[h]
\includegraphics[width=0.4\textwidth]{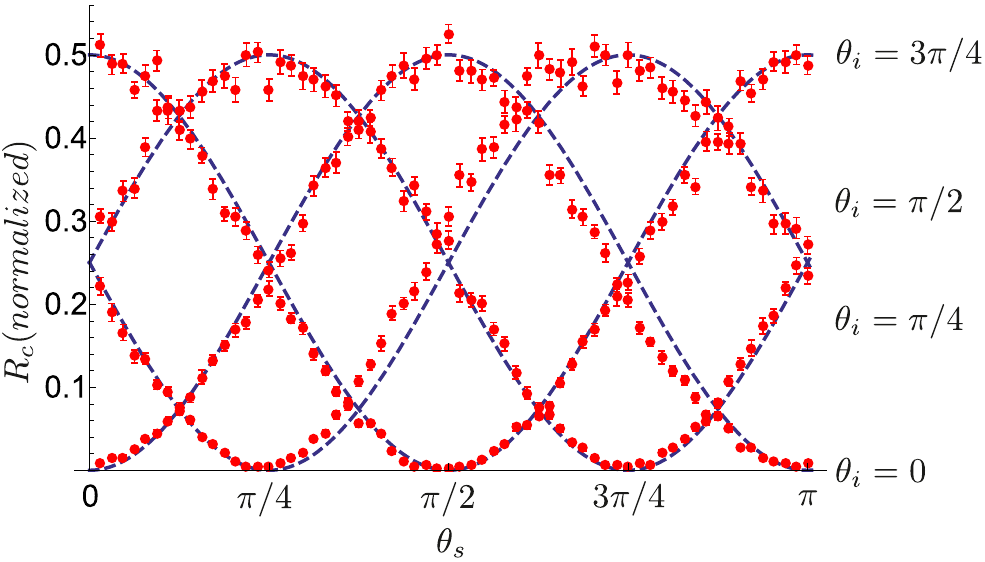}
\parbox{80mm}{\caption{The normalized coincidence count rate $R_{c}$ as a function of the measurement states ``orientaion'' angle $\theta_{s}$ in the signal channel for different angles $\theta_{i}=\{0;~\pi/4;~\pi/2;~3\pi/4\}$ in the idler channel.
}
\label{BellC}}
\end{figure}

In the case of the $\mathrm{HG}_{10}$ pump (Fig.~\ref{fig:hist}b) the two-photon state is indeed very close to the desired spatial Bell state (\ref{Bell}), which we confirmed by performing full state tomography, as well as by violating the Clauser-Horne-Shimony-Holt (CHSH) inequality. 

We performed a Bell-type inequality experiment in a way similar to the one used in \cite{Padgett_PRA10} for the OAM modes. The holograms displayed on SLM2 in the signal and idler channels corresponded to projections on the states 
$\ket{\theta_{s,i}}=\cos (\theta_{s,i}/2) \ket{\mathrm{HG}_{00}}+\sin (\theta_{s,i}/2) \ket{\mathrm{HG}_{10}}.
$
So the coincidence counting rate is $R_{c}(\theta_{s},\theta_{i}) \propto |\bra{\theta_{s}} \langle \theta_{i} \ket{\Psi^{+}}|^{2} \propto \sin(\theta_{s}-\theta_{i})$. The experimentally obtained dependencies are shown in Fig.~\ref{BellC}. The value of the CHSH parameter $S$, which may be calculated from these data, has to be less than 2 for any local realistic theory. We have clearly violated this inequality with the experimentally obtained value of $S=2.81 \pm 0.05$. 

In order to reconstruct the full state density matrix, we performed the full state tomography in the subspaces of higher dimensions: $d=3\times3$ and $d=6\times6$ (see Supplementary Materials for details). 
\begin{figure}[h]
	\center{\includegraphics[width=0.95\linewidth]{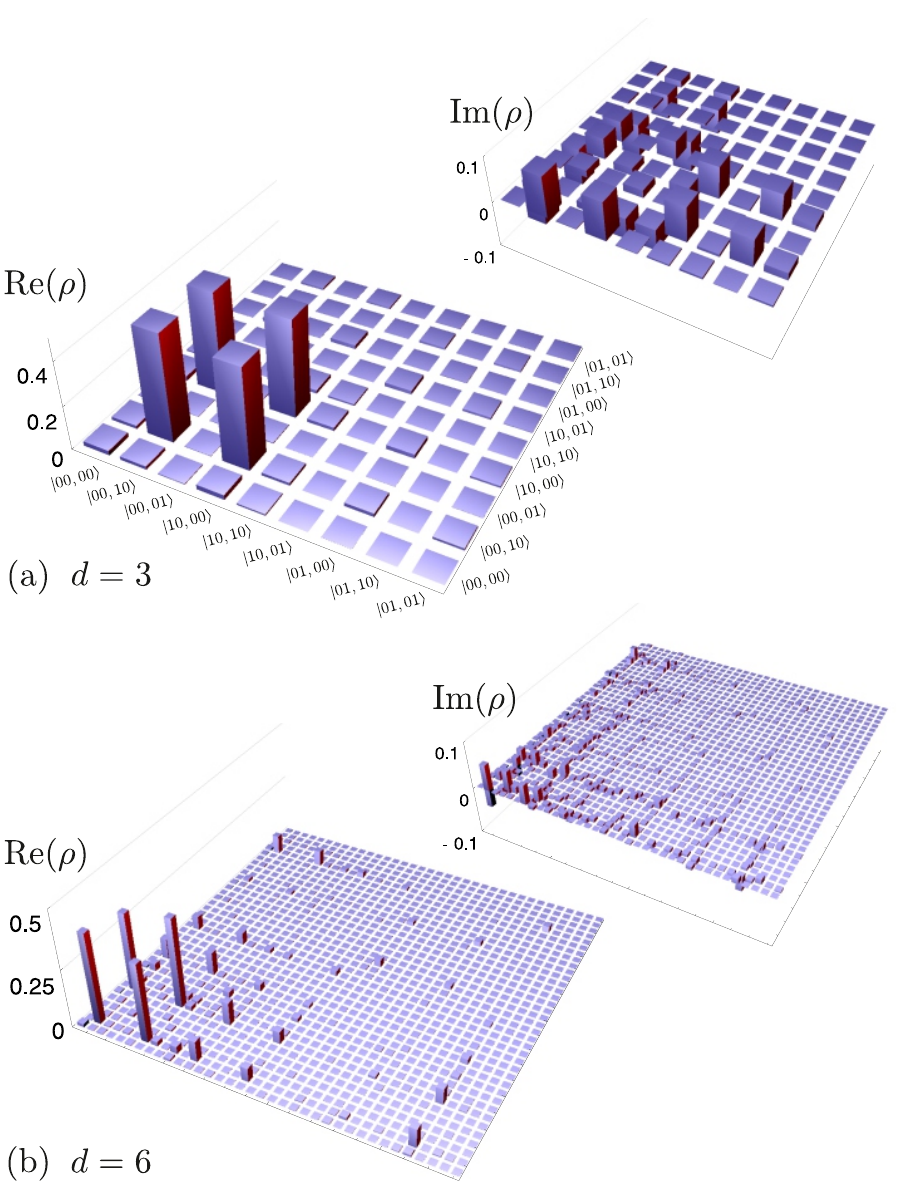}} 
\caption{The density matrix $\rho$ of the generated Bell state: (a) reconstructed in the $d=3\times3$ subspace and (b) reconstructed in the $d=6\times6$ subspace. Here the labels $\ket{jk,ut}$ denote  $\ket{\mathrm{HG}_{jk},\mathrm{HG}_{ut}}$.
} \label{fig:tomo}
\end{figure}

The real and imaginary parts of the reconstructed density matrix $\rho$ are shown in Fig.~\ref{fig:tomo}(a). The fidelity with the target Bell state $\rho_{th}=\ket{\Psi^{+}} \bra{\Psi^{+}}$, given by $F(\rho,\rho_{th})=[\mathrm{Tr}\sqrt{\sqrt{\rho_{th}}\rho\sqrt{\rho_{th}}}]^{2}$, was found to be $0.97 \pm 0.01$. When the state is reconstructed in a larger subspace (Fig.~\ref{fig:tomo}(b)) the fidelity reduces to $F(\rho,\rho_{th}) = 0.72 \pm 0.02$. This kind of fidelity degradation with the increasing dimension was observed before in the experiments with OAM \cite{Boyd,jack2009}. We attribute such behavior to both the reduction of the statistical sample size and to the alignment errors, which become critical as the structure of the detection modes grows in complexity. The presence of anti-diagonal peaks $|C_{ju}^{(1)}|^{2}$ with $j+u=3$ on the histogram in the Fig.~\ref{fig:hist}(b) due to non-Gaussianity of the two-photon amplitude also reduces the fidelity. 

\emph{Discussion and conclusion.}
The results reported here correspond to the $\ket{\Psi^+}$ state, the $\ket{\Psi^-}$ is readily generated by installing an additional mirror in one of the down-conversion channels. The mirror flips the sign of the $\mathrm{HG_{10}}$ mode, while leaving the Gaussian mode intact, thus performing the required transformation. The generation of positively-correlated $\ket{\Phi^{\pm}}$ states is more tricky, since ideally it requires an SPDC state with exactly two Schmidt modes. This cannot be achieved in a completely postselection-free manner, however, by exploiting the pump in a superposition of $\mathrm{HG_{00}}$ and $\mathrm{HG_{20}}$ modes, one may achieve a probability of success for postselecting $\ket{\Phi^{\pm}}$ as high as 70\% (see Supplementary Materials). At the same time, most of the practical entanglement based communication protocols only require a high-fidelity source of one of the Bell states, which is exactly what our method is tailored to achieve.

One of the main advantages of our source is its brightness. The observed coincidence counting rate per 1~mW of pump power was $R_c\sim 0.1$~kHz/mW. The overall loss in our detection scheme are high with the dominating sources being: the SLM2 with amplitude modulation ($\sim 80\%$), the detection efficiency of the photon counters ($\sim 20\%$), non-unity transmission of the interference filter ($\sim50\%$). Taking that into account, we can estimate the pair generation rate as $R\sim15$~kHz/mW, which is comparable with the rates for the sources of polarization entangled pairs using similar crystals \cite{fedrizzi2007}.

The quality of the Bell states produced by the source is mainly limited by the residual population in the $n+m\geq2$ subspace, which is a consequence of non-gaussianity of the exact two-photon amplitude. We believe, that the fidelity of the state may be further improved by adaptively tailoring the spatial profile of the pump to maximize the overlap of the down-converted state with the desired one. Besides that, exploiting higher order pump modes together with adaptive control opens the way for engineering highly-correlated, symmetric spatial states of precisely controlled dimensionality which is one of the directions for further work.

In conclusion, we have experimentally demonstrated the bright source of photon pairs in a spatial Bell state. The source exhibits remarkable features: the spatial state is confined to a 4-dimensional subspace with negligible population of higher-order modes; within this subspace the generated state has high fidelity with an ideal Bell state without any postselection. Controlled dimensionality and lack of irrelevant higher-order modes make this source attractive for applications in quantum communication via turbulent free-space channels \cite{Zeilinger_PNAS15} and few-mode fibers \cite{Richardson_NPhoton13} with significant mode-mixing. Thus it is an important step towards practical implementation of quantum communication protocols based on spatial entanglement.  

This work was in supported in part by the European Union Seventh Framework Programme under grant agreement no. 308803 (project BRISQ2) and RFBR grant 16-32-00889.

\bibliographystyle{apsrev4-1}
\bibliography{paper}

\begin{thebibliography}{37}%
\makeatletter
\providecommand \@ifxundefined [1]{%
 \@ifx{#1\undefined}
}%
\providecommand \@ifnum [1]{%
 \ifnum #1\expandafter \@firstoftwo
 \else \expandafter \@secondoftwo
 \fi
}%
\providecommand \@ifx [1]{%
 \ifx #1\expandafter \@firstoftwo
 \else \expandafter \@secondoftwo
 \fi
}%
\providecommand \natexlab [1]{#1}%
\providecommand \enquote  [1]{``#1''}%
\providecommand \bibnamefont  [1]{#1}%
\providecommand \bibfnamefont [1]{#1}%
\providecommand \citenamefont [1]{#1}%
\providecommand \href@noop [0]{\@secondoftwo}%
\providecommand \href [0]{\begingroup \@sanitize@url \@href}%
\providecommand \@href[1]{\@@startlink{#1}\@@href}%
\providecommand \@@href[1]{\endgroup#1\@@endlink}%
\providecommand \@sanitize@url [0]{\catcode `\\12\catcode `\$12\catcode
  `\&12\catcode `\#12\catcode `\^12\catcode `\_12\catcode `\%12\relax}%
\providecommand \@@startlink[1]{}%
\providecommand \@@endlink[0]{}%
\providecommand \url  [0]{\begingroup\@sanitize@url \@url }%
\providecommand \@url [1]{\endgroup\@href {#1}{\urlprefix }}%
\providecommand \urlprefix  [0]{URL }%
\providecommand \Eprint [0]{\href }%
\providecommand \doibase [0]{http://dx.doi.org/}%
\providecommand \selectlanguage [0]{\@gobble}%
\providecommand \bibinfo  [0]{\@secondoftwo}%
\providecommand \bibfield  [0]{\@secondoftwo}%
\providecommand \translation [1]{[#1]}%
\providecommand \BibitemOpen [0]{}%
\providecommand \bibitemStop [0]{}%
\providecommand \bibitemNoStop [0]{.\EOS\space}%
\providecommand \EOS [0]{\spacefactor3000\relax}%
\providecommand \BibitemShut  [1]{\csname bibitem#1\endcsname}%
\let\auto@bib@innerbib\@empty
\bibitem [{\citenamefont {Ekert}(1991)}]{Ekert}%
  \BibitemOpen
  \bibfield  {author} {\bibinfo {author} {\bibfnamefont {A.~K.}\ \bibnamefont
  {Ekert}},\ }\href {http://dx.doi.org/10.1103/PhysRevLett.67.661} {\bibfield
  {journal} {\bibinfo  {journal} {Phys. Rev. Lett.}\ }\textbf {\bibinfo
  {volume} {67}},\ \bibinfo {pages} {661} (\bibinfo {year} {1991})}\BibitemShut
  {NoStop}%
\bibitem [{\citenamefont {Ac\'{\i}n}\ \emph {et~al.}(2006)\citenamefont
  {Ac\'{\i}n}, \citenamefont {Gisin},\ and\ \citenamefont
  {Masanes}}]{Acin_PRL06}%
  \BibitemOpen
  \bibfield  {author} {\bibinfo {author} {\bibfnamefont {A.}~\bibnamefont
  {Ac\'{\i}n}}, \bibinfo {author} {\bibfnamefont {N.}~\bibnamefont {Gisin}}, \
  and\ \bibinfo {author} {\bibfnamefont {L.}~\bibnamefont {Masanes}},\ }\href
  {\doibase 10.1103/PhysRevLett.97.120405} {\bibfield  {journal} {\bibinfo
  {journal} {Phys. Rev. Lett.}\ }\textbf {\bibinfo {volume} {97}},\ \bibinfo
  {pages} {120405} (\bibinfo {year} {2006})}\BibitemShut {NoStop}%
\bibitem [{\citenamefont {Kwiat}\ \emph {et~al.}(1995)\citenamefont {Kwiat},
  \citenamefont {Mattle}, \citenamefont {Weinfurter}, \citenamefont
  {Zeilinger}, \citenamefont {Sergienko},\ and\ \citenamefont
  {Shih}}]{kwiat1995}%
  \BibitemOpen
  \bibfield  {author} {\bibinfo {author} {\bibfnamefont {P.~G.}\ \bibnamefont
  {Kwiat}}, \bibinfo {author} {\bibfnamefont {K.}~\bibnamefont {Mattle}},
  \bibinfo {author} {\bibfnamefont {H.}~\bibnamefont {Weinfurter}}, \bibinfo
  {author} {\bibfnamefont {A.}~\bibnamefont {Zeilinger}}, \bibinfo {author}
  {\bibfnamefont {A.~V.}\ \bibnamefont {Sergienko}}, \ and\ \bibinfo {author}
  {\bibfnamefont {Y.}~\bibnamefont {Shih}},\ }\href
  {http://dx.doi.org/10.1103/PhysRevLett.75.4337} {\bibfield  {journal}
  {\bibinfo  {journal} {Phys. Rev. Lett.}\ }\textbf {\bibinfo {volume} {75}},\
  \bibinfo {pages} {4337} (\bibinfo {year} {1995})}\BibitemShut {NoStop}%
\bibitem [{\citenamefont {Mair}\ \emph {et~al.}(2001)\citenamefont {Mair},
  \citenamefont {Vaziri}, \citenamefont {Weihs},\ and\ \citenamefont
  {Zeilinger}}]{mair2001}%
  \BibitemOpen
  \bibfield  {author} {\bibinfo {author} {\bibfnamefont {A.}~\bibnamefont
  {Mair}}, \bibinfo {author} {\bibfnamefont {A.}~\bibnamefont {Vaziri}},
  \bibinfo {author} {\bibfnamefont {G.}~\bibnamefont {Weihs}}, \ and\ \bibinfo
  {author} {\bibfnamefont {A.}~\bibnamefont {Zeilinger}},\ }\href
  {http://www.nature.com/nature/journal/v412/n6844/abs/412313a0.html}
  {\bibfield  {journal} {\bibinfo  {journal} {Nature}\ }\textbf {\bibinfo
  {volume} {412}},\ \bibinfo {pages} {313} (\bibinfo {year}
  {2001})}\BibitemShut {NoStop}%
\bibitem [{\citenamefont {Yarnall}\ \emph {et~al.}(2007)\citenamefont
  {Yarnall}, \citenamefont {Abouraddy}, \citenamefont {Saleh},\ and\
  \citenamefont {Teich}}]{Teich_PRL2007}%
  \BibitemOpen
  \bibfield  {author} {\bibinfo {author} {\bibfnamefont {T.}~\bibnamefont
  {Yarnall}}, \bibinfo {author} {\bibfnamefont {A.~F.}\ \bibnamefont
  {Abouraddy}}, \bibinfo {author} {\bibfnamefont {B.~E.}\ \bibnamefont
  {Saleh}}, \ and\ \bibinfo {author} {\bibfnamefont {M.~C.}\ \bibnamefont
  {Teich}},\ }\href
  {http://journals.aps.org/prl/abstract/10.1103/PhysRevLett.99.170408}
  {\bibfield  {journal} {\bibinfo  {journal} {Phys. Rev. Lett.}\ }\textbf
  {\bibinfo {volume} {99}},\ \bibinfo {pages} {170408} (\bibinfo {year}
  {2007})}\BibitemShut {NoStop}%
\bibitem [{\citenamefont {Jack}\ \emph {et~al.}(2010)\citenamefont {Jack},
  \citenamefont {Yao}, \citenamefont {Leach}, \citenamefont {Romero},
  \citenamefont {Franke-Arnold}, \citenamefont {Ireland}, \citenamefont
  {Barnett},\ and\ \citenamefont {Padgett}}]{Padgett_PRA10}%
  \BibitemOpen
  \bibfield  {author} {\bibinfo {author} {\bibfnamefont {B.}~\bibnamefont
  {Jack}}, \bibinfo {author} {\bibfnamefont {A.}~\bibnamefont {Yao}}, \bibinfo
  {author} {\bibfnamefont {J.}~\bibnamefont {Leach}}, \bibinfo {author}
  {\bibfnamefont {J.}~\bibnamefont {Romero}}, \bibinfo {author} {\bibfnamefont
  {S.}~\bibnamefont {Franke-Arnold}}, \bibinfo {author} {\bibfnamefont
  {D.}~\bibnamefont {Ireland}}, \bibinfo {author} {\bibfnamefont
  {S.}~\bibnamefont {Barnett}}, \ and\ \bibinfo {author} {\bibfnamefont
  {M.}~\bibnamefont {Padgett}},\ }\href
  {http://dx.doi.org/10.1103/PhysRevA.81.043844} {\bibfield  {journal}
  {\bibinfo  {journal} {Phys. Rev. A}\ }\textbf {\bibinfo {volume} {81}},\
  \bibinfo {pages} {043844} (\bibinfo {year} {2010})}\BibitemShut {NoStop}%
\bibitem [{\citenamefont {D'Ambrosio}\ \emph {et~al.}(2012)\citenamefont
  {D'Ambrosio}, \citenamefont {Nagali}, \citenamefont {Walborn}, \citenamefont
  {Aolita}, \citenamefont {Slussarenko}, \citenamefont {Marrucci},\ and\
  \citenamefont {Sciarrino}}]{Sciarrino_NComm12}%
  \BibitemOpen
  \bibfield  {author} {\bibinfo {author} {\bibfnamefont {V.}~\bibnamefont
  {D'Ambrosio}}, \bibinfo {author} {\bibfnamefont {E.}~\bibnamefont {Nagali}},
  \bibinfo {author} {\bibfnamefont {S.~P.}\ \bibnamefont {Walborn}}, \bibinfo
  {author} {\bibfnamefont {L.}~\bibnamefont {Aolita}}, \bibinfo {author}
  {\bibfnamefont {S.}~\bibnamefont {Slussarenko}}, \bibinfo {author}
  {\bibfnamefont {L.}~\bibnamefont {Marrucci}}, \ and\ \bibinfo {author}
  {\bibfnamefont {F.}~\bibnamefont {Sciarrino}},\ }\href
  {http://dx.doi.org/10.1038/ncomms1951} {\bibfield  {journal} {\bibinfo
  {journal} {Nature communications}\ }\textbf {\bibinfo {volume} {3}},\
  \bibinfo {pages} {961} (\bibinfo {year} {2012})}\BibitemShut {NoStop}%
\bibitem [{\citenamefont {McLaren}\ \emph {et~al.}(2014)\citenamefont
  {McLaren}, \citenamefont {Mhlanga}, \citenamefont {Padgett}, \citenamefont
  {Roux},\ and\ \citenamefont {Forbes}}]{mclaren2014}%
  \BibitemOpen
  \bibfield  {author} {\bibinfo {author} {\bibfnamefont {M.}~\bibnamefont
  {McLaren}}, \bibinfo {author} {\bibfnamefont {T.}~\bibnamefont {Mhlanga}},
  \bibinfo {author} {\bibfnamefont {M.~J.}\ \bibnamefont {Padgett}}, \bibinfo
  {author} {\bibfnamefont {F.~S.}\ \bibnamefont {Roux}}, \ and\ \bibinfo
  {author} {\bibfnamefont {A.}~\bibnamefont {Forbes}},\ }\href
  {http://www.nature.com/ncomms/2014/140206/ncomms4248/full/ncomms4248.html}
  {\bibfield  {journal} {\bibinfo  {journal} {Nature communications}\ }\textbf
  {\bibinfo {volume} {5}} (\bibinfo {year} {2014})}\BibitemShut {NoStop}%
\bibitem [{\citenamefont {Barreiro}\ \emph {et~al.}(2005)\citenamefont
  {Barreiro}, \citenamefont {Langford}, \citenamefont {Peters},\ and\
  \citenamefont {Kwiat}}]{barreiro2005}%
  \BibitemOpen
  \bibfield  {author} {\bibinfo {author} {\bibfnamefont {J.~T.}\ \bibnamefont
  {Barreiro}}, \bibinfo {author} {\bibfnamefont {N.~K.}\ \bibnamefont
  {Langford}}, \bibinfo {author} {\bibfnamefont {N.~A.}\ \bibnamefont
  {Peters}}, \ and\ \bibinfo {author} {\bibfnamefont {P.~G.}\ \bibnamefont
  {Kwiat}},\ }\href {http://dx.doi.org/10.1103/PhysRevLett.95.260501}
  {\bibfield  {journal} {\bibinfo  {journal} {Phys. Rev. Lett.}\ }\textbf
  {\bibinfo {volume} {95}},\ \bibinfo {pages} {260501} (\bibinfo {year}
  {2005})}\BibitemShut {NoStop}%
\bibitem [{\citenamefont {Leach}\ \emph {et~al.}(2009)\citenamefont {Leach},
  \citenamefont {Jack}, \citenamefont {Romero}, \citenamefont {Ritsch-Marte},
  \citenamefont {Boyd}, \citenamefont {Jha}, \citenamefont {Barnett},
  \citenamefont {Franke-Arnold},\ and\ \citenamefont
  {Padgett}}]{Padgett_OpEx2009}%
  \BibitemOpen
  \bibfield  {author} {\bibinfo {author} {\bibfnamefont {J.}~\bibnamefont
  {Leach}}, \bibinfo {author} {\bibfnamefont {B.}~\bibnamefont {Jack}},
  \bibinfo {author} {\bibfnamefont {J.}~\bibnamefont {Romero}}, \bibinfo
  {author} {\bibfnamefont {M.}~\bibnamefont {Ritsch-Marte}}, \bibinfo {author}
  {\bibfnamefont {R.}~\bibnamefont {Boyd}}, \bibinfo {author} {\bibfnamefont
  {A.}~\bibnamefont {Jha}}, \bibinfo {author} {\bibfnamefont {S.}~\bibnamefont
  {Barnett}}, \bibinfo {author} {\bibfnamefont {S.}~\bibnamefont
  {Franke-Arnold}}, \ and\ \bibinfo {author} {\bibfnamefont {M.}~\bibnamefont
  {Padgett}},\ }\href {https://doi.org/10.1364/OE.17.008287} {\bibfield
  {journal} {\bibinfo  {journal} {Optics express}\ }\textbf {\bibinfo {volume}
  {17}},\ \bibinfo {pages} {8287} (\bibinfo {year} {2009})}\BibitemShut
  {NoStop}%
\bibitem [{\citenamefont {Torres}\ \emph {et~al.}(2003)\citenamefont {Torres},
  \citenamefont {Alexandrescu},\ and\ \citenamefont {Torner}}]{Torres_PRA2003}%
  \BibitemOpen
  \bibfield  {author} {\bibinfo {author} {\bibfnamefont {J.}~\bibnamefont
  {Torres}}, \bibinfo {author} {\bibfnamefont {A.}~\bibnamefont
  {Alexandrescu}}, \ and\ \bibinfo {author} {\bibfnamefont {L.}~\bibnamefont
  {Torner}},\ }\href {https://doi.org/10.1103/PhysRevA.68.050301} {\bibfield
  {journal} {\bibinfo  {journal} {Physical Review A}\ }\textbf {\bibinfo
  {volume} {68}},\ \bibinfo {pages} {050301} (\bibinfo {year}
  {2003})}\BibitemShut {NoStop}%
\bibitem [{\citenamefont {Agnew}\ \emph {et~al.}(2013)\citenamefont {Agnew},
  \citenamefont {Salvail}, \citenamefont {Leach},\ and\ \citenamefont
  {Boyd}}]{Boyd_PRL2013}%
  \BibitemOpen
  \bibfield  {author} {\bibinfo {author} {\bibfnamefont {M.}~\bibnamefont
  {Agnew}}, \bibinfo {author} {\bibfnamefont {J.~Z.}\ \bibnamefont {Salvail}},
  \bibinfo {author} {\bibfnamefont {J.}~\bibnamefont {Leach}}, \ and\ \bibinfo
  {author} {\bibfnamefont {R.~W.}\ \bibnamefont {Boyd}},\ }\href {\doibase
  10.1103/PhysRevLett.111.030402} {\bibfield  {journal} {\bibinfo  {journal}
  {Phys. Rev. Lett.}\ }\textbf {\bibinfo {volume} {111}},\ \bibinfo {pages}
  {030402} (\bibinfo {year} {2013})}\BibitemShut {NoStop}%
\bibitem [{\citenamefont {Abouraddy}\ \emph {et~al.}(2007)\citenamefont
  {Abouraddy}, \citenamefont {Yarnall}, \citenamefont {Saleh},\ and\
  \citenamefont {Teich}}]{Teich_PRA2007}%
  \BibitemOpen
  \bibfield  {author} {\bibinfo {author} {\bibfnamefont {A.~F.}\ \bibnamefont
  {Abouraddy}}, \bibinfo {author} {\bibfnamefont {T.}~\bibnamefont {Yarnall}},
  \bibinfo {author} {\bibfnamefont {B.~E.~A.}\ \bibnamefont {Saleh}}, \ and\
  \bibinfo {author} {\bibfnamefont {M.~C.}\ \bibnamefont {Teich}},\ }\href
  {\doibase 10.1103/PhysRevA.75.052114} {\bibfield  {journal} {\bibinfo
  {journal} {Phys. Rev. A}\ }\textbf {\bibinfo {volume} {75}},\ \bibinfo
  {pages} {052114} (\bibinfo {year} {2007})}\BibitemShut {NoStop}%
\bibitem [{\citenamefont {Romero}\ \emph {et~al.}(2013)\citenamefont {Romero},
  \citenamefont {Giovannini}, \citenamefont {Tasca}, \citenamefont {Barnett},\
  and\ \citenamefont {Padgett}}]{tailored}%
  \BibitemOpen
  \bibfield  {author} {\bibinfo {author} {\bibfnamefont {J.}~\bibnamefont
  {Romero}}, \bibinfo {author} {\bibfnamefont {D.}~\bibnamefont {Giovannini}},
  \bibinfo {author} {\bibfnamefont {D.}~\bibnamefont {Tasca}}, \bibinfo
  {author} {\bibfnamefont {S.}~\bibnamefont {Barnett}}, \ and\ \bibinfo
  {author} {\bibfnamefont {M.}~\bibnamefont {Padgett}},\ }\href
  {http://iopscience.iop.org/1367-2630/15/8/083047} {\bibfield  {journal}
  {\bibinfo  {journal} {New Journal of Physics}\ }\textbf {\bibinfo {volume}
  {15}},\ \bibinfo {pages} {083047} (\bibinfo {year} {2013})}\BibitemShut
  {NoStop}%
\bibitem [{\citenamefont {Burlakov}\ \emph {et~al.}(1997)\citenamefont
  {Burlakov}, \citenamefont {Chekhova}, \citenamefont {Klyshko}, \citenamefont
  {Kulik}, \citenamefont {Penin}, \citenamefont {Shih},\ and\ \citenamefont
  {Strekalov}}]{Burlakov_PRA97}%
  \BibitemOpen
  \bibfield  {author} {\bibinfo {author} {\bibfnamefont {A.~V.}\ \bibnamefont
  {Burlakov}}, \bibinfo {author} {\bibfnamefont {M.~V.}\ \bibnamefont
  {Chekhova}}, \bibinfo {author} {\bibfnamefont {D.~N.}\ \bibnamefont
  {Klyshko}}, \bibinfo {author} {\bibfnamefont {S.~P.}\ \bibnamefont {Kulik}},
  \bibinfo {author} {\bibfnamefont {A.~N.}\ \bibnamefont {Penin}}, \bibinfo
  {author} {\bibfnamefont {Y.~H.}\ \bibnamefont {Shih}}, \ and\ \bibinfo
  {author} {\bibfnamefont {D.~V.}\ \bibnamefont {Strekalov}},\ }\href {\doibase
  10.1103/PhysRevA.56.3214} {\bibfield  {journal} {\bibinfo  {journal} {Phys.
  Rev. A}\ }\textbf {\bibinfo {volume} {56}},\ \bibinfo {pages} {3214}
  (\bibinfo {year} {1997})}\BibitemShut {NoStop}%
\bibitem [{\citenamefont {Monken}\ \emph {et~al.}(1998)\citenamefont {Monken},
  \citenamefont {Ribeiro},\ and\ \citenamefont {P{\'a}dua}}]{monken1998}%
  \BibitemOpen
  \bibfield  {author} {\bibinfo {author} {\bibfnamefont {C.~H.}\ \bibnamefont
  {Monken}}, \bibinfo {author} {\bibfnamefont {P.~S.}\ \bibnamefont {Ribeiro}},
  \ and\ \bibinfo {author} {\bibfnamefont {S.}~\bibnamefont {P{\'a}dua}},\
  }\href {http://dx.doi.org/10.1103/PhysRevA.57.3123} {\bibfield  {journal}
  {\bibinfo  {journal} {Phys. Rev. A}\ }\textbf {\bibinfo {volume} {57}},\
  \bibinfo {pages} {3123} (\bibinfo {year} {1998})}\BibitemShut {NoStop}%
\bibitem [{\citenamefont {Walborn}\ \emph {et~al.}(2004)\citenamefont
  {Walborn}, \citenamefont {De~Oliveira}, \citenamefont {Thebaldi},\ and\
  \citenamefont {Monken}}]{walborn2004}%
  \BibitemOpen
  \bibfield  {author} {\bibinfo {author} {\bibfnamefont {S.}~\bibnamefont
  {Walborn}}, \bibinfo {author} {\bibfnamefont {A.}~\bibnamefont
  {De~Oliveira}}, \bibinfo {author} {\bibfnamefont {R.}~\bibnamefont
  {Thebaldi}}, \ and\ \bibinfo {author} {\bibfnamefont {C.}~\bibnamefont
  {Monken}},\ }\href {http://dx.doi.org/10.1103/PhysRevA.69.023811} {\bibfield
  {journal} {\bibinfo  {journal} {Phys. Rev. A}\ }\textbf {\bibinfo {volume}
  {69}},\ \bibinfo {pages} {023811} (\bibinfo {year} {2004})}\BibitemShut
  {NoStop}%
\bibitem [{\citenamefont {Molina-Terriza}\ \emph {et~al.}(2001)\citenamefont
  {Molina-Terriza}, \citenamefont {Torres},\ and\ \citenamefont
  {Torner}}]{Torner_PRL01}%
  \BibitemOpen
  \bibfield  {author} {\bibinfo {author} {\bibfnamefont {G.}~\bibnamefont
  {Molina-Terriza}}, \bibinfo {author} {\bibfnamefont {J.~P.}\ \bibnamefont
  {Torres}}, \ and\ \bibinfo {author} {\bibfnamefont {L.}~\bibnamefont
  {Torner}},\ }\href {\doibase 10.1103/PhysRevLett.88.013601} {\bibfield
  {journal} {\bibinfo  {journal} {Phys. Rev. Lett.}\ }\textbf {\bibinfo
  {volume} {88}},\ \bibinfo {pages} {013601} (\bibinfo {year}
  {2001})}\BibitemShut {NoStop}%
\bibitem [{\citenamefont {Walborn}\ \emph {et~al.}(2003)\citenamefont
  {Walborn}, \citenamefont {de~Oliveira}, \citenamefont {P\'adua},\ and\
  \citenamefont {Monken}}]{Walborn_PRL03}%
  \BibitemOpen
  \bibfield  {author} {\bibinfo {author} {\bibfnamefont {S.~P.}\ \bibnamefont
  {Walborn}}, \bibinfo {author} {\bibfnamefont {A.~N.}\ \bibnamefont
  {de~Oliveira}}, \bibinfo {author} {\bibfnamefont {S.}~\bibnamefont
  {P\'adua}}, \ and\ \bibinfo {author} {\bibfnamefont {C.~H.}\ \bibnamefont
  {Monken}},\ }\href {\doibase 10.1103/PhysRevLett.90.143601} {\bibfield
  {journal} {\bibinfo  {journal} {Phys. Rev. Lett.}\ }\textbf {\bibinfo
  {volume} {90}},\ \bibinfo {pages} {143601} (\bibinfo {year}
  {2003})}\BibitemShut {NoStop}%
\bibitem [{\citenamefont {Romero}\ \emph {et~al.}(2012)\citenamefont {Romero},
  \citenamefont {Giovannini}, \citenamefont {McLaren}, \citenamefont {Galvez},
  \citenamefont {Forbes},\ and\ \citenamefont {Padgett}}]{Padgett_JoO12}%
  \BibitemOpen
  \bibfield  {author} {\bibinfo {author} {\bibfnamefont {J.}~\bibnamefont
  {Romero}}, \bibinfo {author} {\bibfnamefont {D.}~\bibnamefont {Giovannini}},
  \bibinfo {author} {\bibfnamefont {M.~G.}\ \bibnamefont {McLaren}}, \bibinfo
  {author} {\bibfnamefont {E.~J.}\ \bibnamefont {Galvez}}, \bibinfo {author}
  {\bibfnamefont {A.}~\bibnamefont {Forbes}}, \ and\ \bibinfo {author}
  {\bibfnamefont {M.~J.}\ \bibnamefont {Padgett}},\ }\href
  {http://stacks.iop.org/2040-8986/14/i=8/a=085401} {\bibfield  {journal}
  {\bibinfo  {journal} {Journal of Optics}\ }\textbf {\bibinfo {volume} {14}},\
  \bibinfo {pages} {085401} (\bibinfo {year} {2012})}\BibitemShut {NoStop}%
\bibitem [{\citenamefont {Walborn}\ \emph {et~al.}(2005)\citenamefont
  {Walborn}, \citenamefont {P{\'a}dua},\ and\ \citenamefont
  {Monken}}]{Walborn_PRA2005}%
  \BibitemOpen
  \bibfield  {author} {\bibinfo {author} {\bibfnamefont {S.}~\bibnamefont
  {Walborn}}, \bibinfo {author} {\bibfnamefont {S.}~\bibnamefont {P{\'a}dua}},
  \ and\ \bibinfo {author} {\bibfnamefont {C.}~\bibnamefont {Monken}},\ }\href
  {http://dx.doi.org/10.1103/PhysRevA.71.053812} {\bibfield  {journal}
  {\bibinfo  {journal} {Phys. Rev. A}\ }\textbf {\bibinfo {volume} {71}},\
  \bibinfo {pages} {053812} (\bibinfo {year} {2005})}\BibitemShut {NoStop}%
\bibitem [{\citenamefont {Ren}\ \emph {et~al.}(2005)\citenamefont {Ren},
  \citenamefont {Guo}, \citenamefont {Li},\ and\ \citenamefont
  {Guo}}]{Guo_PhysLettA05}%
  \BibitemOpen
  \bibfield  {author} {\bibinfo {author} {\bibfnamefont {X.-F.}\ \bibnamefont
  {Ren}}, \bibinfo {author} {\bibfnamefont {G.-P.}\ \bibnamefont {Guo}},
  \bibinfo {author} {\bibfnamefont {J.}~\bibnamefont {Li}}, \ and\ \bibinfo
  {author} {\bibfnamefont {G.-C.}\ \bibnamefont {Guo}},\ }\href {\doibase
  http://dx.doi.org/10.1016/j.physleta.2005.04.060} {\bibfield  {journal}
  {\bibinfo  {journal} {Physics Letters A}\ }\textbf {\bibinfo {volume}
  {341}},\ \bibinfo {pages} {81 } (\bibinfo {year} {2005})}\BibitemShut
  {NoStop}%
\bibitem [{\citenamefont {Walborn}\ and\ \citenamefont
  {Pimentel}(2012)}]{Walborn_JPhysB12}%
  \BibitemOpen
  \bibfield  {author} {\bibinfo {author} {\bibfnamefont {S.}~\bibnamefont
  {Walborn}}\ and\ \bibinfo {author} {\bibfnamefont {A.}~\bibnamefont
  {Pimentel}},\ }\href {http://iopscience.iop.org/0953-4075/45/16/165502}
  {\bibfield  {journal} {\bibinfo  {journal} {Journal of Physics B: Atomic,
  Molecular and Optical Physics}\ }\textbf {\bibinfo {volume} {45}},\ \bibinfo
  {pages} {165502} (\bibinfo {year} {2012})}\BibitemShut {NoStop}%
\bibitem [{\citenamefont {Law}\ and\ \citenamefont {Eberly}(2004)}]{Eberly}%
  \BibitemOpen
  \bibfield  {author} {\bibinfo {author} {\bibfnamefont {C.~K.}\ \bibnamefont
  {Law}}\ and\ \bibinfo {author} {\bibfnamefont {J.~H.}\ \bibnamefont
  {Eberly}},\ }\href {\doibase 10.1103/PhysRevLett.92.127903} {\bibfield
  {journal} {\bibinfo  {journal} {Phys. Rev. Lett.}\ }\textbf {\bibinfo
  {volume} {92}},\ \bibinfo {pages} {127903} (\bibinfo {year}
  {2004})}\BibitemShut {NoStop}%
\bibitem [{\citenamefont {Straupe}\ \emph {et~al.}(2011)\citenamefont
  {Straupe}, \citenamefont {Ivanov}, \citenamefont {Kalinkin}, \citenamefont
  {Bobrov},\ and\ \citenamefont {Kulik}}]{Straupe}%
  \BibitemOpen
  \bibfield  {author} {\bibinfo {author} {\bibfnamefont {S.}~\bibnamefont
  {Straupe}}, \bibinfo {author} {\bibfnamefont {D.}~\bibnamefont {Ivanov}},
  \bibinfo {author} {\bibfnamefont {A.}~\bibnamefont {Kalinkin}}, \bibinfo
  {author} {\bibfnamefont {I.}~\bibnamefont {Bobrov}}, \ and\ \bibinfo {author}
  {\bibfnamefont {S.}~\bibnamefont {Kulik}},\ }\href {\doibase
  10.1103/PhysRevA.83.060302} {\bibfield  {journal} {\bibinfo  {journal} {Phys.
  Rev. A}\ }\textbf {\bibinfo {volume} {83}},\ \bibinfo {pages} {060302}
  (\bibinfo {year} {2011})}\BibitemShut {NoStop}%
\bibitem [{\citenamefont {Bolduc}\ \emph {et~al.}(2013)\citenamefont {Bolduc},
  \citenamefont {Bent}, \citenamefont {Santamato}, \citenamefont {Karimi},\
  and\ \citenamefont {Boyd}}]{Exact}%
  \BibitemOpen
  \bibfield  {author} {\bibinfo {author} {\bibfnamefont {E.}~\bibnamefont
  {Bolduc}}, \bibinfo {author} {\bibfnamefont {N.}~\bibnamefont {Bent}},
  \bibinfo {author} {\bibfnamefont {E.}~\bibnamefont {Santamato}}, \bibinfo
  {author} {\bibfnamefont {E.}~\bibnamefont {Karimi}}, \ and\ \bibinfo {author}
  {\bibfnamefont {R.}~\bibnamefont {Boyd}},\ }\href
  {http://dx.doi.org/10.1364/OL.38.003546} {\bibfield  {journal} {\bibinfo
  {journal} {Optics letters}\ }\textbf {\bibinfo {volume} {38}},\ \bibinfo
  {pages} {3546} (\bibinfo {year} {2013})}\BibitemShut {NoStop}%
\bibitem [{\citenamefont {Bent}\ \emph {et~al.}(2015)\citenamefont {Bent},
  \citenamefont {Qassim}, \citenamefont {Tahir}, \citenamefont {Sych},
  \citenamefont {Leuchs}, \citenamefont {S{\'a}nchez-Soto}, \citenamefont
  {Karimi},\ and\ \citenamefont {Boyd}}]{bent2015}%
  \BibitemOpen
  \bibfield  {author} {\bibinfo {author} {\bibfnamefont {N.}~\bibnamefont
  {Bent}}, \bibinfo {author} {\bibfnamefont {H.}~\bibnamefont {Qassim}},
  \bibinfo {author} {\bibfnamefont {A.}~\bibnamefont {Tahir}}, \bibinfo
  {author} {\bibfnamefont {D.}~\bibnamefont {Sych}}, \bibinfo {author}
  {\bibfnamefont {G.}~\bibnamefont {Leuchs}}, \bibinfo {author} {\bibfnamefont
  {L.~L.}\ \bibnamefont {S{\'a}nchez-Soto}}, \bibinfo {author} {\bibfnamefont
  {E.}~\bibnamefont {Karimi}}, \ and\ \bibinfo {author} {\bibfnamefont
  {R.}~\bibnamefont {Boyd}},\ }\href
  {https://doi.org/10.1103/PhysRevX.5.041006} {\bibfield  {journal} {\bibinfo
  {journal} {Phys. Rev. X}\ }\textbf {\bibinfo {volume} {5}},\ \bibinfo {pages}
  {041006} (\bibinfo {year} {2015})}\BibitemShut {NoStop}%
\bibitem [{\citenamefont {Bobrov}\ \emph {et~al.}(2015)\citenamefont {Bobrov},
  \citenamefont {Kovlakov}, \citenamefont {Markov}, \citenamefont {Straupe},\
  and\ \citenamefont {Kulik}}]{bobrov2015}%
  \BibitemOpen
  \bibfield  {author} {\bibinfo {author} {\bibfnamefont {I.}~\bibnamefont
  {Bobrov}}, \bibinfo {author} {\bibfnamefont {E.}~\bibnamefont {Kovlakov}},
  \bibinfo {author} {\bibfnamefont {A.}~\bibnamefont {Markov}}, \bibinfo
  {author} {\bibfnamefont {S.}~\bibnamefont {Straupe}}, \ and\ \bibinfo
  {author} {\bibfnamefont {S.}~\bibnamefont {Kulik}},\ }\href
  {http://dx.doi.org/10.1364/OE.23.000649} {\bibfield  {journal} {\bibinfo
  {journal} {Optics express}\ }\textbf {\bibinfo {volume} {23}},\ \bibinfo
  {pages} {649} (\bibinfo {year} {2015})}\BibitemShut {NoStop}%
\bibitem [{Note1()}]{Note1}%
  \BibitemOpen
  \bibinfo {note} {We assume the Gaussian pump beam profile $\protect \mathcal
  {E}_{p}(x,w_{p}) \propto \protect \qopname \relax
  o{exp}{(-x^{2}/w_{p}^{2})}$, and define the waist to be $w_p$. We should
  note, that the width parameter $w$ used in (\ref {SchmidtN}) is the beam
  waist times $\protect \sqrt {2}$, to comply with the definitions in \cite
  {Eberly} and the following literature}\BibitemShut {NoStop}%
\bibitem [{\citenamefont {Salakhutdinov}\ \emph {et~al.}(2012)\citenamefont
  {Salakhutdinov}, \citenamefont {Eliel},\ and\ \citenamefont
  {L{\"o}ffler}}]{salakhutdinov2012}%
  \BibitemOpen
  \bibfield  {author} {\bibinfo {author} {\bibfnamefont {V.}~\bibnamefont
  {Salakhutdinov}}, \bibinfo {author} {\bibfnamefont {E.}~\bibnamefont
  {Eliel}}, \ and\ \bibinfo {author} {\bibfnamefont {W.}~\bibnamefont
  {L{\"o}ffler}},\ }\href {http://dx.doi.org/10.1103/PhysRevLett.108.173604}
  {\bibfield  {journal} {\bibinfo  {journal} {Phys. Rev. Lett.}\ }\textbf
  {\bibinfo {volume} {108}},\ \bibinfo {pages} {173604} (\bibinfo {year}
  {2012})}\BibitemShut {NoStop}%
\bibitem [{\citenamefont {Miatto}\ \emph {et~al.}(2012)\citenamefont {Miatto},
  \citenamefont {Pires}, \citenamefont {Barnett},\ and\ \citenamefont {van
  Exter}}]{Miatto_EPJD2012}%
  \BibitemOpen
  \bibfield  {author} {\bibinfo {author} {\bibfnamefont {F.~M.}\ \bibnamefont
  {Miatto}}, \bibinfo {author} {\bibfnamefont {H.~D.~L.}\ \bibnamefont
  {Pires}}, \bibinfo {author} {\bibfnamefont {S.~M.}\ \bibnamefont {Barnett}},
  \ and\ \bibinfo {author} {\bibfnamefont {M.~P.}\ \bibnamefont {van Exter}},\
  }\href {https://doi.org/10.1140/epjd/e2012-30035-3} {\bibfield  {journal}
  {\bibinfo  {journal} {The European Physical Journal D}\ }\textbf {\bibinfo
  {volume} {66}},\ \bibinfo {pages} {1} (\bibinfo {year} {2012})}\BibitemShut
  {NoStop}%
\bibitem [{\citenamefont {Agnew}\ \emph {et~al.}(2011)\citenamefont {Agnew},
  \citenamefont {Leach}, \citenamefont {McLaren}, \citenamefont {Roux},\ and\
  \citenamefont {Boyd}}]{Boyd}%
  \BibitemOpen
  \bibfield  {author} {\bibinfo {author} {\bibfnamefont {M.}~\bibnamefont
  {Agnew}}, \bibinfo {author} {\bibfnamefont {J.}~\bibnamefont {Leach}},
  \bibinfo {author} {\bibfnamefont {M.}~\bibnamefont {McLaren}}, \bibinfo
  {author} {\bibfnamefont {F.~S.}\ \bibnamefont {Roux}}, \ and\ \bibinfo
  {author} {\bibfnamefont {R.~W.}\ \bibnamefont {Boyd}},\ }\href
  {http://dx.doi.org/10.1103/PhysRevA.84.062101} {\bibfield  {journal}
  {\bibinfo  {journal} {Phys. Rev. A}\ }\textbf {\bibinfo {volume} {84}},\
  \bibinfo {pages} {062101} (\bibinfo {year} {2011})}\BibitemShut {NoStop}%
\bibitem [{\citenamefont {Jack}\ \emph {et~al.}(2009)\citenamefont {Jack},
  \citenamefont {Leach}, \citenamefont {Ritsch}, \citenamefont {Barnett},
  \citenamefont {Padgett},\ and\ \citenamefont {Franke-Arnold}}]{jack2009}%
  \BibitemOpen
  \bibfield  {author} {\bibinfo {author} {\bibfnamefont {B.}~\bibnamefont
  {Jack}}, \bibinfo {author} {\bibfnamefont {J.}~\bibnamefont {Leach}},
  \bibinfo {author} {\bibfnamefont {H.}~\bibnamefont {Ritsch}}, \bibinfo
  {author} {\bibfnamefont {S.}~\bibnamefont {Barnett}}, \bibinfo {author}
  {\bibfnamefont {M.}~\bibnamefont {Padgett}}, \ and\ \bibinfo {author}
  {\bibfnamefont {S.}~\bibnamefont {Franke-Arnold}},\ }\href
  {http://dx.doi.org/10.1088/1367-2630/11/10/103024} {\bibfield  {journal}
  {\bibinfo  {journal} {New Journal of Physics}\ }\textbf {\bibinfo {volume}
  {11}},\ \bibinfo {pages} {103024} (\bibinfo {year} {2009})}\BibitemShut
  {NoStop}%
\bibitem [{\citenamefont {Fedrizzi}\ \emph {et~al.}(2007)\citenamefont
  {Fedrizzi}, \citenamefont {Herbst}, \citenamefont {Poppe}, \citenamefont
  {Jennewein},\ and\ \citenamefont {Zeilinger}}]{fedrizzi2007}%
  \BibitemOpen
  \bibfield  {author} {\bibinfo {author} {\bibfnamefont {A.}~\bibnamefont
  {Fedrizzi}}, \bibinfo {author} {\bibfnamefont {T.}~\bibnamefont {Herbst}},
  \bibinfo {author} {\bibfnamefont {A.}~\bibnamefont {Poppe}}, \bibinfo
  {author} {\bibfnamefont {T.}~\bibnamefont {Jennewein}}, \ and\ \bibinfo
  {author} {\bibfnamefont {A.}~\bibnamefont {Zeilinger}},\ }\href
  {http://dx.doi.org/10.1364/OE.15.015377} {\bibfield  {journal} {\bibinfo
  {journal} {Optics Express}\ }\textbf {\bibinfo {volume} {15}},\ \bibinfo
  {pages} {15377} (\bibinfo {year} {2007})}\BibitemShut {NoStop}%
\bibitem [{\citenamefont {Krenn}\ \emph {et~al.}(2015)\citenamefont {Krenn},
  \citenamefont {Handsteiner}, \citenamefont {Fink}, \citenamefont {Fickler},\
  and\ \citenamefont {Zeilinger}}]{Zeilinger_PNAS15}%
  \BibitemOpen
  \bibfield  {author} {\bibinfo {author} {\bibfnamefont {M.}~\bibnamefont
  {Krenn}}, \bibinfo {author} {\bibfnamefont {J.}~\bibnamefont {Handsteiner}},
  \bibinfo {author} {\bibfnamefont {M.}~\bibnamefont {Fink}}, \bibinfo {author}
  {\bibfnamefont {R.}~\bibnamefont {Fickler}}, \ and\ \bibinfo {author}
  {\bibfnamefont {A.}~\bibnamefont {Zeilinger}},\ }\href {\doibase
  10.1073/pnas.1517574112} {\bibfield  {journal} {\bibinfo  {journal}
  {Proceedings of the National Academy of Sciences}\ }\textbf {\bibinfo
  {volume} {112}},\ \bibinfo {pages} {14197} (\bibinfo {year}
  {2015})}\BibitemShut {NoStop}%
\bibitem [{\citenamefont {Richardson}\ \emph {et~al.}(2013)\citenamefont
  {Richardson}, \citenamefont {Fini},\ and\ \citenamefont
  {Nelson}}]{Richardson_NPhoton13}%
  \BibitemOpen
  \bibfield  {author} {\bibinfo {author} {\bibfnamefont {D.}~\bibnamefont
  {Richardson}}, \bibinfo {author} {\bibfnamefont {J.}~\bibnamefont {Fini}}, \
  and\ \bibinfo {author} {\bibfnamefont {L.}~\bibnamefont {Nelson}},\ }\href
  {http://dx.doi.org/10.1038/nphoton.2013.94} {\bibfield  {journal} {\bibinfo
  {journal} {Nature Photonics}\ }\textbf {\bibinfo {volume} {7}},\ \bibinfo
  {pages} {354} (\bibinfo {year} {2013})}\BibitemShut {NoStop}%
\bibitem [{\citenamefont {James}\ \emph {et~al.}(2001)\citenamefont {James},
  \citenamefont {Kwiat}, \citenamefont {Munro},\ and\ \citenamefont
  {White}}]{KwiatQ}%
  \BibitemOpen
  \bibfield  {author} {\bibinfo {author} {\bibfnamefont {D.~F.}\ \bibnamefont
  {James}}, \bibinfo {author} {\bibfnamefont {P.~G.}\ \bibnamefont {Kwiat}},
  \bibinfo {author} {\bibfnamefont {W.~J.}\ \bibnamefont {Munro}}, \ and\
  \bibinfo {author} {\bibfnamefont {A.~G.}\ \bibnamefont {White}},\ }\href
  {http://dx.doi.org/10.1103/PhysRevA.64.052312} {\bibfield  {journal}
  {\bibinfo  {journal} {Phys. Rev. A}\ }\textbf {\bibinfo {volume} {64}},\
  \bibinfo {pages} {052312} (\bibinfo {year} {2001})}\BibitemShut {NoStop}%
\end{thebibliography}%

\newpage

\onecolumngrid

\section*{\large Supplementary Information}
\twocolumngrid

\section*{Schmidt number measurements}

\begin{figure}[h]
	\includegraphics[width=0.9\linewidth]{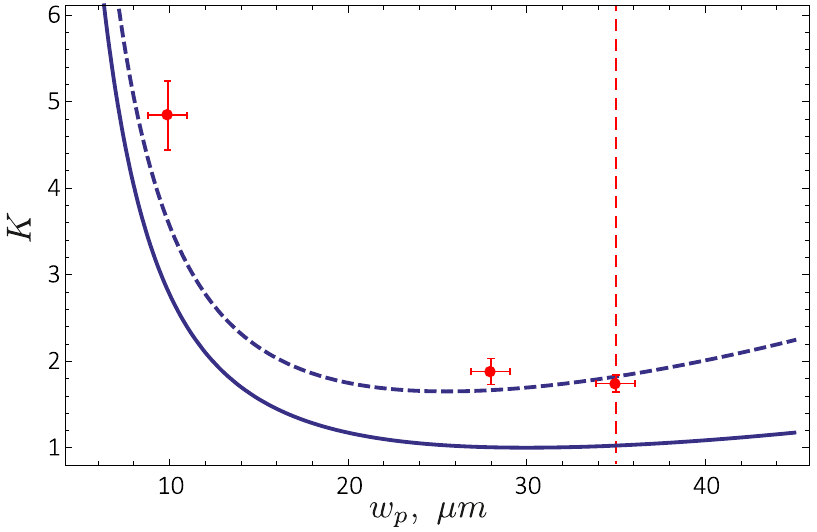}
	\caption{The dependence of the Schmidt number $K$ on the pump beam waist $w_p$. Red dots correspond to the values, estimated from the experimentally measured histograms; solid blue line -- to the prediction of the double-Gaussian approximation; dashed blue line -- to the prediction of the $(\alpha,\beta)$-modification from \cite{Miatto_EPJD2012}. The red dashed vertical line marks the value of the pump beam waist for which all the results in the main article were obtained.}
	\label{fig:Schmidt_curve} 
\end{figure}

At the first stage of the experiment, our main goal was to achieve the lowest possible value of the Schmidt number. As was mentioned above, the analytic expression for the Schmidt number $K$
\begin{equation}
	\label{SchmidtN2}
	K=\Bigl(\dfrac{w^{2}+\delta^{2}}{2w\delta} \Bigr)^{2}
\end{equation}
as a function of the pump beam waist $w$ is obtained by the substitution:
\begin{equation}
\label{Amplitude_Suppl}
\sinc (\delta^{2} (\vec{k}_{s_\bot}-\vec{k}_{i_\bot})^{2})\rightarrow \exp (-\delta^{2} (\vec{k}_{s_\bot}-\vec{k}_{i_\bot})^{2})
\end{equation}
in the following expression for the biphoton amplitude:
\begin{equation}
\label{Amplitude_Suppl}
\Psi(\vec{k_{s \perp}},\vec{k_{i \perp}})\propto \mathcal{E}_{p}^{(nm)}(\vec{k_{s \perp}}+\vec{k_{i \perp}}) \sinc \Bigr(\dfrac{L(\vec{k}_{s_\bot}-\vec{k}_{i_\bot})^{2}}{4k_{p}} \Bigl).
\end{equation}
According to the numerical results in \cite{Miatto_EPJD2012}, the equation (\ref{SchmidtN2}) could be improved with additional factors $\alpha$ and $\beta$
\begin{equation}
	\label{SchmidtN2M}
	K=\beta \Bigl(\dfrac{w^{2}+\alpha^{2} \delta^{2}}{2w \alpha \delta} \Bigr)^{2},
\end{equation}
where $\alpha = 0.85$ and $\beta =1.65$. Fig.~\ref{fig:Schmidt_curve} depicts the Schmidt number $K$ as a function of the pump beam waist $w_{p}$ for two cases approaches: the solid line corresponds to the equation (\ref{SchmidtN2}) and the dashed line represents its modification (\ref{SchmidtN2M}). From this plot it is easy to see that the minimum of the modified curve rises to the value of $K=1.65$ and is shifted to the lower value of $w_{p}=25.5~\mu m$. This result is in a good agreement with the experimentally obtained value of $w_{p}\simeq 25~ \mu m$ corresponding to the maximal efficiency of photon pair coupling to single-mode fibers, as previously reported in \cite{fedrizzi2007}. 

We tried the configuration of lenses L1 with focal length 150 mm and L2 with focal length 100 mm, corresponding to $w_{p}=28.0 \pm 1.0~ \mu m$ in our experiment. Surprisingly, we observed the increase of the Schmidt number estimate up to $K=1.88 \pm 0.15$.  This discrepancy might be related to the optical aberrations and beam waists misalignment along the pump beam propagation direction in the nonlinear crystal. In addition, we performed a Schmidt number measurement for $w_{p}=9.9 \pm 1.1~ \mu m$. All experimental results are shown in Fig.~\ref{fig:Schmidt_curve} as red points. 

We need to note again, that the Schmidt number estimation method mentioned in the main article is not direct, because we used only absolute values of experimentally obtained coefficients $C^{(0)}_{aa}$  instead of full reconstruction of all coefficients $C^{(0)}_{jukt}$ in the decomposition. However, our numerical simulation showed that this kind of estimation error is still less than the statistical errors.

\section*{State tomography}

An estimate for the density matrix was obtained by minimizing the $\chi^2$ quantity for the predicted probabilities $p^{(P)}_{i}$ and the measured probabilities $p^{(M)}_{i}\propto R_{c}$ (the coincidence rates $R_{c}$ were normalized to sum to unity):

\begin{equation}
\chi^{2}=\sum^{N}_{i=1} \dfrac{\left(p^{(M)}_{i}-p^{(P)}_{i}\right)^{2}}{p^{(P)}_{i}}.
\end{equation}

For the reconstruction in a $3\times 3$ dimensional subspace, we describe an estimate for a density matrix by a Cholesky decomposition of a $9\times 9$ matrix with 80 independent real parameters. Thus the reconstructed matrix is Hermitian and positive semidefinite with unit trace by construction \cite{KwiatQ}. We also chose an over-complete set of measurements to increase the accuracy of the reconstruction. The state was projected onto a set of $N=15 \times 15$  eigenvectors of the generalized Gell-Mann matrices with the modes $\ket{\mathrm{HG}_{00}}$, $\ket{\mathrm{HG}_{10}}$ and $\ket{\mathrm{HG}_{01}}$ as a measurement basis.

In the case of $d=6\times6$ subspace, we enriched the measurement basis of our tomography protocol by adding $\ket{\mathrm{HG}_{20}}$, $\ket{\mathrm{HG}_{11}}$ and $\ket{\mathrm{HG}_{02}}$ modes. This time we used a set of only $N=36 \times 36$ projectors due to the complexity of the reconstruction process rapidly increasing with $d$. 

\section*{Correlated Bell states}

\begin{figure}[h] 
	\includegraphics[width=0.95\linewidth]{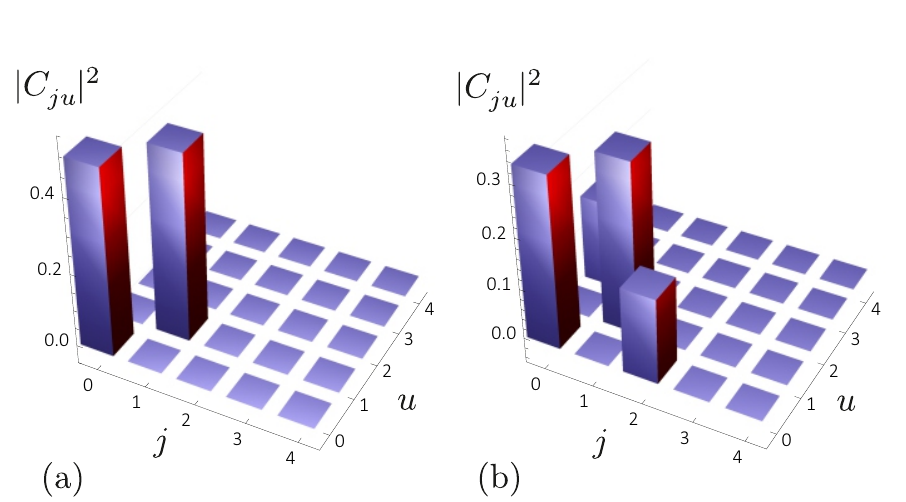}
	
	\caption{Coefficients $|C_{ju}|^{2}$ for an ideal correlated Bell states $\ket{\Phi^{(\pm)}}$ (a), and the results of the numerical calculation for the SPDC state with a pump in a superposition of $\mathrm{HG}_{00}$ and $\mathrm{HG}_{20}$ modes (b).} 
	\label{fig:correlated_Bell}
\end{figure}

Our method may be straightforwardly used to generate anti-correlated Bell states $$\ket{\Psi^\pm}= \frac{1}{\sqrt{2}}\left(\ket{\mathrm{HG}_{00},\mathrm{HG}_{10}}\pm\ket{\mathrm{HG}_{10},\mathrm{HG}_{00}}\right).$$ Generation of correlated states 
$$\ket{\Phi^\pm}= \frac{1}{\sqrt{2}}\left(\ket{\mathrm{HG}_{00},\mathrm{HG}_{00}}\pm\ket{\mathrm{HG}_{10},\mathrm{HG}_{10}}\right)$$
requires an SPDC state with exactly two non-zero Schmidt eigenvalues, as shown in Fig.~\ref{fig:correlated_Bell}(a). That is impossible to achieve in a post-selection free manner. However, under a double-Gaussian approximation one can generate a state shown in Fig.~\ref{fig:correlated_Bell}(b) by using the pump beam in a superposition of a Gaussian and $\mathrm{HG}_{20}$ modes: $\mathcal{E}_{p}(k_{x},k_{y},w) =(\alpha \mathrm{HG}_{0}(k_{x},w)+\sqrt{1-\alpha^{2}}\mathrm{HG}_{2}(k_{x},w)) \times \mathrm{HG}_{0}(k_{y},w)$ with $\alpha=0.895$. The probability of successfully post-selecting the correlated Bell state in this case is given by the overlap of the generated state with an ideal Bell state, shown in Fig.~\ref{fig:correlated_Bell}(a). This overlap (fidelity) may be calculated to be $F=\left|\langle \Psi | \Phi^{+}\rangle \right|^2 = 0.67$. The $\ket{\Phi^-}$ state may also be generated by simply inverting the sign of $\alpha$ in the pump modal superposition. So potentially our scheme is capable of producing the correlated Bell states with an upper bound of success probability of about 70\%. 

\end{document}